\documentstyle[12pt,aaspp4]{article}
\lefthead{Das and Chakrabarti}
\righthead{Outflows from Accretion Disks}

\def\lsim{\lower.5ex\hbox{$\; \buildrel < \over \sim \;$}}
\def\gsim{\lower.5ex\hbox{$\; \buildrel > \over \sim \;$}}
\begin{document}

\title{Mass Outflow Rate From Advective Accretion Disks around Compact Objects}

\author{Tapas K. Das and Sandip K.\ Chakrabarti} 

\affil{\small S.N. Bose National Centre for Basic Sciences,\\
JD Block, Salt Lake, Sector-III, Calcutta-700091\\
e-mail: tdas@boson.bose.res.in \& chakraba@boson.bose.res.in}

\begin{abstract}

We compute mass outflow rates from advective accretion disks around compact
objects, such as neutron stars and black holes. These computations,
for the first time, are done using combinations of exact transonic inflow and outflow
solutions which may or may not form standing shock waves. Assuming that
the bulk of the outflow is from the effective boundary layers of these objects, 
we find that the ratio of the outflow rate and inflow rate varies 
anywhere from a few percent to even close to a hundred percent
(i.e., close to disk evacuation case) depending on the initial parameters of the 
disk, the degree of compression of matter near the centrifugal barrier, 
and the polytropic index of the flow. Our result, in general, 
matches  with the outflow rates obtained 
through a fully time-dependent numerical simulation. In some
region of the parameter space when the standing shock does not form, 
our results {\it indicate} that the disk may be evacuated
and may produce quiescence states. 

\end{abstract}
 

\noindent Submitted to ApJ in April 1998; Still trying to teach the referee the meaning
of centrifugal barrier and the funnel wall.

\section{Introduction}

One of the signatures of activities around compact objects is the presence
of jets and outflows. Outflows carry away angular momentum from the
accretion disk and are partially responsible for the accretion itself.
Active galaxies and quasars are supposed to harbor black holes at their centers and at the
same time produce cosmic radio jets through which immense amount
of matter and energy are ejected out of the core of the galaxies
(See, Begelman, Blandford \& Rees, 1984 and Chakrabarti, 1996a for reviews).
Similarly, micro-quasars have also been discovered very recently
where outflows are formed from stellar black hole candidates
(Mirabel \& Rodriguez, 1994). Many of these outflows
show superluminal motions which are probably due to magnetic effects. With hydrodynamic
effects alone, which we are employing in this paper, one should not be able to 
accelerate the flow more than the initial sound velocity. A well known stellar object
SS433 (which is believed to be a neutron star), where pure hydrodynamic effects
may be operating, produces outflows with a speed roughly one-third of the speed of light.

There are several models in the literature which study the origin,
formation and collimation of these outflows. Difference
between stellar outflows and outflows from
these systems is that the outflows in these systems have to
form out of the inflowing material only. This is because black holes
and neutron stars have no atmospheres of their own. The models
present in the literature are roughly of three types. The first type of
solutions confine themselves to the jet properties only, completely
decoupled from the internal properties of accretion disks. They
study the effects of hydrodynamic or magneto-hydrodynamic
pressures on the origin of jets (e.g., Blandford \& Payne 1982;
Ferrari et al., 1985; Fukue 1987; Contopoulus 1985; Chakrabarti, 1990).
In the second type, efforts are made
to correlate the internal disk structure with that of the
outflow using both hydrodynamic (e.g., Chakrabarti 1986)
and magnetohydrodynamic considerations (K\"onigl 1989; Chakrabarti \&
Bhaskaran 1992). In the third type, numerical simulations are carried out
to actually see how matter is deflected from the equatorial plane 
towards the axis (e.g., Hawley, Smarr \& Wilson, 1984; 
Eggum, Katz \& Coroniti, 1985; Molteni, Lanzafame \& Chakrabarti, 1994; 
Molteni, Ryu \& Chakrabarti, 1996; Ryu, Chakrabarti \& Molteni, 1997). 
From the analytical front, although the wind type solutions and accretion type solutions
come out of the same set of governing equations (Chakrabarti 1990), there were no attempt
to find connections among them. As a result, the estimation of the
outflow rate from the inflow rate has always been impossible.
On the other hand, the mass outflow rate of the normal stars are 
calculated very accurately from the stellar luminosity. Theory of 
radiatively driven winds seems to be very well understood 
(e.g., Castor, Abott \& Klein, 1975). The simplicity of black holes
and neutron stars lie in the fact that they do not have atmospheres.
But the disks surrounding them have, and similar method as employed
in stellar atmospheres should be applicable to the disks.
Our approach in this paper is precisely this. We first
determine the properties of the rotating inflow and outflow
and identify solutions to connect them. In this manner we
self-consistently determine the mass outflow rates.

Before we present our results, we describe basic properties of
the rotating inflow and outflow. A rotating inflow with a specific angular momentum $\lambda (x)$
entering into a black hole  will have angular momentum $\lambda \sim $  constant 
close to the black hole for any moderate viscous stress. This is because the
viscous time scale is generally much longer compared to the 
infall time scale and even though at the outer edges the angular momentum distribution
may be Keplerian or even super-Keplerian matter would be highly sub-Keplerian close to the
black hole. This is because the flow {\it has to enter} through the horizon with velocity of light
and presence of this large inertial (ram) force, in addition to usual gravitational
and centrifugal forces, makes the flow sub-Keplerian. Almost constant angular momentum 
produces a very strong centrifugal force $\lambda^2/x^3$ which 
increases much faster compared to the gravitational force
$\sim GM/x^2$ and becomes comparable at around $x\sim l^2/GM$,
or, $r_{cb}\sim 2\lambda^2$ where $r$ and $\lambda$ are $x$ and $l$,
written in units of $R_g=2GM/c^2$ and $R_g c = 2GM/c$ respectively. 
The subscript $cb$ under $r$ stands for the centrifugal barrier. (In the
rest of the paper, we use $R_g$ as the length unit, $c$ is the 
unit of velocity, and the mass of the black hole $M$ to the 
unit of mass.) Here, (actually, a little farther out, due to 
thermal pressure) matter starts 
piling up and produces the centrifugal pressure supported boundary layer
(CENBOL). Further close to the black hole, the gravity always wins 
and matter enters the horizon supersonically after passing 
through a sonic point. CENBOL may or may not have a sharp boundary,
depending on whether standing shocks form or not. Generally speaking,
in a polytropic flow, if the polytropic index $\gamma > 1.5$, then 
shocks do not form and if $\gamma <1.5$, only a region of the parameter
space forms the shock (C96b). In any case, the CENBOL forms. In this region 
the flow becomes hotter and denser and for all practical purposes 
behaves as the stellar atmosphere so far as the formation of 
outflows are concerned. Inflows on neutron stars behave similarly,
except that the `hard-surface' inner boundary condition dictates that the
flow remain subsonic between the CENBOL and the surface rather than becoming 
supersonic as in the case of a black hole. In case where the shock does not form, 
regions around pressure maximum achieved just outside the inner sonic 
point would also drive the flow outwards. In the back of our 
mind,  we have the picture of the outflow as obtained by Hawley, 
Smarr \& Wilson (1984), Chakrabarti (1986) and Molteni, Ryu and 
Chakrabarti (1996), namely, that the outflow is thermally and centrifugally
accelerated but confined by external pressure of the ambient medium. 

There are two surfaces of utmost importance in flows with angular momentum. One is the
`funnel wall' where the effective potential (sum of gravitational potential
and the specific rotational energy) vanishes. In the case of a purely rotating 
flow, this is the `zero pressure' surface. Flows {\it cannot} enter inside the
funnel wall because the pressure would be negative. The other surface 
is called the `centrifugal barrier'. This is the surface
where the radial pressure gradient of a purely rotating flow vanishes and is located
{\it outside} the funnel wall simply because the flow pressure is higher than zero
on this surface. Flow with inertial pressure easily crosses this `barrier' and either enters 
into a black hole or flows out as winds depending on its initial parameters (detail
classification of the parameter space is in Chakrabarti, 1989)
In numerical simulations (e.g., Hawley, Smarr \& Wilson, 1984; Molteni, Ryu and Chakrabarti, 1996) 
it is observed that the outflow generally hugs the `funnel wall' and goes out in between
these two surfaces. In this paper we assume precisely this. However, we believe that the
the cross section area could be weakly influenced by the pressence of the small
outward radial motion of the flow. We have ignored such a correction.

Outflow rates from accretion disks around  black hole and neutron 
stars must be related to the properties of CENBOL which in turn, 
depend on the inflow parameters. Subsonic outflows originating from CENBOL 
would pass through sonic points and reach far distances as in wind 
solutions. Assuming free-falling conical polytropic inflow and isothermal
outflows (as in stellar winds), it is easy to estimate the
ratio of outflowing and inflowing rates (Chakrabarti, 1997a; 1998):
$$
\frac{{\dot M}_{out}}{{\dot M}_{in}} = R_{\dot m}=
\frac{\Theta_{out}}{\Theta_{in}}\frac{R}{4} e^{ -(f_0 - \frac{3}{2})} f_0^{3/2}
$$
where, $\Theta_{out}$ and $\Theta_{in}$ are the solid angles of
the outflow and inflow respectively, and 
$$
f_0 = \frac{(2n+1)R}{2n}.
$$
Here, $R$ is the compression ratio of inflowing matter at the CENBOL
and $n=1/(\gamma-1)$ is the polytropic constant. When $\Theta_{out} \sim \Theta_{in}$, 
$R_{\dot m} \sim 0.052$ and $0.266$ for $\gamma=4/3$ and $5/3$ respectively.
Assuming a thin inflow and outflow of $10^o$ conical angle, the
ratio $R_{\dot m}$ becomes  $0.0045$ and $0.023$ respectively. 

The aim of the present paper is to compute the mass loss rate
more realistically than what has been attempted so far.
We calculate this rate as a function of the inflow parameters, such
as specific  energy and angular momentum, accretion rate, polytropic index etc.
We explore both the polytropic and the isothermal outflows. Our conclusions show 
that the outflow rate is sensitive to the specific energy and 
accretion rate of the inflow. Specifically, when the outflow 
is not isothermal, outflow rate generally increases with the specific energy
and the polytropic index $\gamma_{o}$ of the outflow, generally decreases with the
polytropic index $\gamma$ of the inflow, but somewhat insensitive to
the specific angular momentum $\lambda$. 
In the case of isothermal outflow, however, mass loss rate is sensitive to the
inflow rate, since the inflow rate decides the proton temperature of
the advective region of the disk which in turn fixes the outflow temperature. In this
case the outflow is at least partially temperature driven. The outflow rate is 
also found to be anti-correlated with specific angular momentum $\lambda$ of the
flow.

The plan of this paper is the following: In the next Section, we describe our model
and present the governing equations for the inflow and outflow. 
In \S 3, we present the solution procedure of the equations.
In \S 4, we  present results of our computations. 
Finally, in \S 5, we draw our conclusions.

\section{Model Description and Governing Equations}

\subsection{Inflow Model}

For the sake of computation of the inflow
quantities, we assume that the inflow is axisymmetric and thin: $h(r) << r$, 
so that the transverse velocity component could be ignored compared to the 
radial and azimuthal velocity components. We consider
polytropic inflows in vertical equilibrium (otherwise known as 1.5 dimensional
flows, see, Chakrabarti, 1989; hereafter referred to as C89). We 
ignore the self-gravity of the flow and viscosity is assumed to be significant only
at the shock, if present. We do the calculations using Paczy\'nski-Wiita (1980)
potential which mimics surroundings of the Schwarzschild black hole. 
The equations (in dimensionless units) governing the inflow are:\\
\noindent (a) Conservation of specific energy is given by,
$$
{\cal E}=\frac{{u_e}^2}{2} +n{{a_e}^2}+\frac{{\lambda}^2}{2{r^2}}-\frac{1}{2(r-1)} .
\eqno{(1)}
$$
where, $u_e$ and $a_e$ are the radial and polytropic sound 
velocities respectively. $a_e=(\gamma p_e/\rho_e)^{1/2}$,
$p_e$ and $\rho_e$ are the pressure and density of the 
flow. For a polytropic flow, $p_e = K \rho_e^\gamma$,
where $K$ is a constant and is a measure of 
entropy of the flow. Here, $\lambda$ is the specific 
angular momentum and $n$ is the polytropic constant of the inflow 
$n=(\gamma-1)^{-1}$, $\gamma$ is the polytropic index.
The subscript $e$ refers that the quantities 
are measured on the equatorial plane.

Mass conservation equation, apart from a geometric constant, is given by,\\
$$
{{\dot M}_{in}}={u_e}{{\rho}_e}r{h_e}(r),
\eqno{(2)}
$$
where ${h_e}(r)$ is the half-thickness of the flow at radial co-ordinate $r$
having the following expression
$$
{h_e}(r)={a_e}{r^{\frac{1}{2}}}(r-1)\sqrt{\frac{2}{\gamma }}.
\eqno{(3a)}
$$
Another useful way of writing the mass inflow is to introduce an entropy dependent quantity ${\dot{\cal M}}
\propto \gamma^n K^n {\dot M} $  which can be expressed as
$$
{\dot{\cal M}}={u_e}{{a_e}^{\alpha}}{r^{\frac{3}{2}}}(r-1)\sqrt{\frac{2}{\gamma }}
\eqno{(3b)}
$$
Where, ${\dot{\cal M}}$ is really the entropy accretion rate (C89). When the shock
is not present, ${\dot{\cal M}}$ remains constant in a polytropic flow. 
When the shock is present, ${\dot{\cal M}}$ will increase at the shock 
due to increase of entropy. $\alpha=(\gamma+1)/(\gamma-1)=2n+1$. If the 
centrifugal pressure supported shock is present, the usual Rankine-Hugoniot conditions,
namely, conservations of mass, energy and momentum fluxes across the shock 
are to be taken into account (C89) in determining the shock locations. In presence of 
mass loss one must incorporate this effect in the shock condition (see, eq. 15
below).

\subsection{Outflow Models}

We consider two types of outflows. In ordinary stellar mass loss computations
(e.g. Tarafdar, 1988, and references therein), the outflow is assumed to be
isothermal till the sonic point. This assumption is probably justified,
since copious photons from the stellar atmosphere deposit momenta on the
slowly outgoing and expanding outflow and possibly make the flow close to isothermal.
This need not be the case for outflows from compact sources. Centrifugal
pressure supported boundary layers close to the black hole are very hot 
(close to the virial temperature) and most of the photons emitted may be
swallowed by the black holes themselves instead of coming out of the 
region and depositing momentum onto the outflow. Thus, the outflows could be
cooler than isothermal flows. In our first model, we choose polytropic outflows
with same energy as the inflow (i.e., no energy dissipation between the
inflow and outflow) but with a different polytropic index $\gamma_{o} <\gamma$.
Nevertheless, it may be advisable to study the isothermal outflow to find out the
behavior of the extreme case. Thus an isothermal outflow is chosen in our second model.
In each case, of course, we include the possibility that the {\it inflow}
may or may not have standing shocks.

It is to be noted that we chose the outflows to be 
propagating through the funnel wall even though the inflow
is assumed to be thin. This may not be completely self-inconsistent.
On the one hand, our assumption of thin inflow is for the
sake of computation of the thermodynamic quantities only, but the
flow itself need not be physically thin. Secondly, the funnel wall
and the centrifugal barrier are purely geometric surfaces, and they exist
anyway and the outflow could be supported even by ambient medium
which may not necessarily be a part of the disk itself. So, we believe that
our assumptions are not entirely unjustified.

\subsubsection{Polytropic Outflow}

In this case, the energy conservation equation takes the form:
$$
{\cal E}=\frac{\vartheta^2}{2}+{n^\prime}{{a_e}^2}+\frac{{\lambda}^2}{2{{r_m}^2(r)}}
-\frac{1}{2(r-1)}
\eqno{(4)}
$$
and the mass conservation in the outflow takes the form:
$$
{{\dot M}_{out}}={\rho}{\vartheta}{\cal A}(r).
\eqno{(5)}
$$
Here, $n^\prime=(\gamma_{o}-1)^{-1}$ is the polytropic constant of the outflow.
The difference between  eq. (4) and eq. (1) is that, presently,
the rotational energy term contains 
$$
{r_m}(r)=\frac{{\Re}(r)+R(r)}{2},
\eqno{(6a)}
$$
as the mean {\it axial} distance of the flow. The expression of ${{\Re}(r)}$, 
the local radius of the centrifugal barrier comes from balancing 
the centrifugal force with the gravity (Molteni, Ryu \& Chakrabarti, 1996), i.e.,
$$
\frac{{\lambda}^2}{{\Re}^3(r)}=\frac{{\Re}(r)}{2r{(r-1)^2}}.
\eqno{(6b)}
$$
We thus obtain,
$$
{\Re}(r)={\left[ {2{\lambda^2}r{(r-1)^2}}\right ]}^{\frac{1}{4}}
\eqno{(7a)}
$$
And the expression for $R(r)$, the local radius of the funnel wall,
comes from vanishing of total effective potential, i.e.,
$$
{{\Omega}_{toteff}(r)}=-\frac{1}{2(r-1)}+\frac{{\lambda^2}}{2{R^2}(r)}=0
$$
$$
R(r)= \lambda \left [{{(r-1)}}\right]^{1/2}
\eqno{(7b)}
$$
The difference between eq. (5) and eq. (2) is that the area functions are different.
Here, $A(r)$ is the area between the centrifugal barrier and the funnel
wall (see introduction for the motivation).
This is computed with the assumption that the outflow is external pressure
supported, i.e., the centrifugal barrier is in pressure
balanced with the ambient medium. Matter, if pushed hard enough, can cross centrifugal barrier
in black hole accretion (the reason why rapidly rotating matter can enter into a
black hole in the first place). An outward thermal force (such as provided by the 
CENBOL) in between the funnel wall
and the centrifugal barrier causes the flow to come out. Thus the 
cross section of the outflow is,
$$
{\cal A}(r)=\pi[{\Re}^2(r) - R^2 (r)].
\eqno{(8)}
$$
The outflow angular momentum $\lambda$ is chosen to be the same as in the
inflow, i.e., no viscous dissipation is assumed to be present in the inner region of the
flow close to a black hole. Considering that viscous time scales are longer compared to 
the inflow time scale, it may be a good  assumption in the disk, but it may not be
a very good assumption for the outflows which are slow prior to the acceleration and
are therefore, prone to viscous transport of angular momentum. Such detailed study has not been
attempted
here particularly because we know very little about the viscous processes taking place
in the pre-jet flow. Therefore, we concentrate only those cases where the specific angular 
momentum is roughly constant when inflowing matter becomes part of the outflow, although
some estimates of the change in $R_{\dot m}$ is provided when the average angular momentum
of the outflow is lower. Detailed study of the outflow rates in presence of viscosity 
and magnetic field is in progress and would be presented elsewhere.

\subsubsection{Isothermal Outflow}

The integration of the radial momentum equation yields an equation similar to the 
energy equation (eq. 4):
$$
\frac{{\vartheta_{iso}}^2}{2}+{{C_s}^2}ln{\rho}+\frac{\lambda^2}{2{{r_m}(r)}^2}
-\frac{1}{2(r-1)}={\rm Constant}
\eqno{(9)}
$$
In this case the thermal energy term is different, behaving logarithmically. The constant sound speed
of the outflow is $C_s$. The mass conservation equation remains the same:
$$
{{\dot{M}}_{out}}=\rho {\vartheta_{iso}}{{\cal A}(r)}.
\eqno{(10)}
$$
Here, the area function remains the same above. A subscript {\it  iso} of 
velocity ${\vartheta}$ is kept to distinguish from the velocity in 
polytropic case. This is to indicate the 
velocities are measured here using completely different assumptions.

In both the models of the outflow, we assume that the flow is primarily radial. Thus the
$\theta$-component of the velocity is ignored ($\vartheta_\theta  << \vartheta$).

\section{Procedure to solve for disks and outflows simultaneously }

Before we go into the details, a general understanding of the transonic flows 
around a black hole is essential. In Chakrabarti (1989, 1996c), all the solution topologies
of the polytropic flow in pseudo-Newtonian geometry has been provided (e.g., see, Fig. 3 Chakrabarti, 1996c)
In regions {\bf I} and {\bf O} of the parameter space the flow has only one sonic point. Matter with positive
energy at a large distance must pass through that point before entering into the black hole
supersonically. In regions {\bf SA} and {\bf SW} shocks may form in accretion and winds respectively, but no shocks are
expected in winds and accretions if parameters are chosen from these branches. In {\bf NSW} and {\bf NSA},
two saddle type sonic points exist, but no steady shock solutions are possible.

Suppose that matter first enters through the outer sonic point and passes through a shock. At the shock,
part of the incoming matter, having higher entropy density is likely to return back as winds through
a sonic point, other than the one it just entered. Thus a combination of topologies, one from the region
{\bf SA} and the other from the region {\bf O} is required to obtain a full solution. In the absence of the shocks,
the flow is likely to bounce back at the pressure maximum of the inflow and since the outflow would be heated by photons,
and thus have a smaller polytropic constant, the flow would leave the system
through an outer sonic point different from that of the incoming solution. Thus 
finding a complete self-consistent solution boils down to finding the outer sonic point of the
outflow and the mass flux through it. Below we present the list of parameters used in both of our
models and briefly describe the procedure  to obtain a satisfactory solution.

\subsection{Polytropic Outflow}

\noindent We assume that \\
\noindent (a) In this case, a very little amount of total energy is assumed to be lost 
in each bundle of matter as it leaves the disk and joins the jet. 
The specific energy ${\cal E}$ 
remains fixed throughout the flow trajectory as it moves from the disk to the jet. 

\noindent (b) Very little viscosity is present in the flow except at the
place where the shock forms, so that the specific angular momentum $\lambda$
is constant in both inflows and outflows close to the black hole. At the shock, entropy is generated
and hence the outflow is of higher entropy for the same specific energy.

\noindent (c) The polytropic index of the inflow  ($\gamma$) and outflow ($\gamma_{o}$) 
are free parameters and in general, $\gamma_{o} <\gamma$, because of heating effect 
of the outflow (e.g., due to the momentum deposition coming out of the disk surface).
In reality $\gamma_{o}$ is directly related to the heating and cooling processes of the outflow. 
When ${\dot M}_{in}$ is high, heating of outflow by photon momentum deposition is higher, 
and therefore $\gamma_{o} \rightarrow 1$. 

Thus a supply of parameters ${\cal E}$, $\lambda$, $\gamma$ and $\gamma_{o}$ make
a self-consistent computation of $R_{\dot m}$ possible 
when the shock is present. When the shock is absent,
the compression ratio of the gas at the pressure maximum between the inflow and outflow
$R_{comp}$ is supplied as a free parameter, since it may be otherwise very difficult to 
compute satisfactorily. In the presence of shocks, such problems do not arise as the
compression ratio is obtained self-consistently.

\noindent The following procedure is adopted to obtain a complete solution:\\

\noindent (a) From eqs. (1) and (2) we derive an expression for the derivative,
$$
\frac{du}{dr}=\frac{\displaystyle{\frac{\lambda^2}{r^3}+\frac{n{a^2}}{\alpha}+\frac{5r-3}
{r(r-1)}-\frac{1}{2{(r-1)^2}}}}{\displaystyle{u-\frac{2na^2}{\alpha u}}} .
\eqno{(11)}
$$
At the sonic point, the numerator and denominator separately vanish, and give rise to the
so-called sonic point conditions:
$$
a_c=\frac{ \displaystyle{ \frac{1} { 2 { ( {r_c} -1) } ^2 } - \frac{\lambda^2}{{r_c}^3} } } 
{ \displaystyle{ \frac{\alpha({r_c}-1){r_c}}{n(5{r_c}-3)}} }
\eqno{(12a)}
$$
$$
{u_c}={\sqrt{\frac{2n}{\alpha}}}{a_c}
\eqno{(12b)}
$$
where, the subscript $c$ represents the quantities at the sonic point. The derivative
of the flow at the sonic point is computed using the L'Hospital's rule. Using fourth
order Runge-Kutta method $\vartheta (r)$ and $a (r)$ are computed along the flow till
the position where the Rankine-Hugoniot condition is satisfied (if shocks form) and from 
there on the sub-sonic branch is integrated for the accretion as usual (C89). With 
the known $\gamma_{o}$, ${\cal E}$ and  $\lambda$, one can compute the location of the 
outflow sonic point from eqs. (4) and (5), 
$$
\frac{d{\vartheta}}{dr}=
\frac{
\displaystyle{
\frac{a^2}{{\cal A}^2(r)}
\frac{d{\cal A}(r)}{dr}
+\frac{\lambda^2}{{r_m}^3(r)}
\frac{d{r_m}(r)}{dr}-\frac{1}{2(r-1)^2}}}
{\displaystyle{
{\vartheta-\frac{a^2}{\vartheta}}
}
}
\eqno{(13)}
$$
from where the sonic point conditions at the outflow sonic point $r_{co}$ obtained are given by,
$$
\frac{a_{co}^2}{{\cal A}_{co}^2(r)} \frac{d {\cal A}(r)}{dr} |_{co}
+{\frac{\lambda^2}{{{r_m}_{co}}^3(r)}}{{\frac{d{r_m}(r)}{dr}}|_{co}}-\frac{1}
{2{({r_{co}}-1)^2}}=0
\eqno{(14a)}
$$
and
$$
{{\vartheta}_{co}}=a_{co} .
\eqno{(14b)}
$$
At the outer sonic point, the derivative of $\vartheta$ is computed using the L'Hospital's rule
and the Runge-Kutta method is used to integrate towards the black hole to compute the velocity
of the outflow at the shock location. The density of the outflow at the shock 
is computed by distributing the post-shock dense matter of the disk into
spherical shell of $4\pi$ solid angle. The outflow rate is then computed using eq. (5).

It is to be noted that when the outflows are produced, one cannot use the usual Rankine-Hugoniot
relations at the shock location, since mass flux is no longer conserved {\it in accretion}, but part of
it is lost in the outflow. Accordingly, we use,
$$
{\dot M}_+ = (1-R_{\dot m}) {\dot M}_- 
\eqno{(15)}
$$
where, the subscripts $+$ and $-$ denote the pre- and post-shock values respectively. Since due to the
loss of matter in the post-shock region, the post-shock pressure goes down, the shock recedes
backward for the same value of incoming energy, angular momentum \& polytropic index. The combination of
three changes, namely, the increase in the cross-sectional area of the outflow and the
launching velocity of the outflow and the decrease in the post-shock density decides whether the 
net outflow rate would increased or decreased than from the case when the 
exact Rankine-Hugoniot relation was used.

In the case where the shocks do not form, the procedure is a bit different. It is assumed that the
maximum amount of matter comes out from the place of the disk where the thermal pressure of the inflow attains 
its maximum. The expression for the polytropic pressure for the inflow in vertical equilibrium is,
$$
{{\cal P}_e}(r)=\frac{{{a_e}^{2(n+1)}}{{\dot M}_{in}}}{{\gamma^{(1+n)}}{\dot{\cal  M}}}
\eqno{(16)}
$$
This is maximized and the outflow is assumed to have the same quasi-conical shape with annular
cross-section ${\cal A} (r)$ between the funnel wall and the centrifugal barrier as already defined.
In the absence of shocks the compression ratio of the flow between the incoming flow and outgoing
flow at the pressure maximum cannot be computed self-consistently unlike the case when the shock
was present. Thus this ratio is chosen freely. We take the guidance for this number from 
what was obtained in the case when shocks are formed. However, in this case even when the
mass loss takes place, {\it the location} of the pressure maximum remains unchanged. Since the
compression ratio  $R_{comp}$ is a free parameter, $R_{\dot m}$ remains unchanged for a given
$R_{comp}$. Let us assume that ${\dot \mu}_{-}$ is the {\it actual} mass inflow rate and it is 
same before and after the pressure maximum had the mass loss rate been negligible. Let ${\dot \mu}_{+}$
be the mass inflow rate {\it after } the pressure maximum, when the loss due to outflow is taken into account.
Then, by definition, ${\dot \mu}_{-}={\dot M}_{out}+{\dot \mu}_{+}$ and $R_{\dot m}={\dot M}_{out}/{\dot \mu}_{-}$.
Thus the {\it actual} ratio of the mass outflow rate and the mass inflow rate, when the mass
loss is taken into consideration is given by,
$$
\frac{\dot M_{out}}{\dot \mu_{+}} = \frac{R_{\dot m}}{1-R_{\dot m}}.
\eqno{(17a)}
$$
However, this static consideration is valid only when $R_{\dot m} <1$. Otherwise, we must have,
$$
-\frac{dM_{disk}}{dt} +{\dot \mu}_- = {\dot \mu}_+ + {\dot M}_{out}
$$
i.e.,
$$
-\frac{dM_{disk}}{dt} = {\dot \mu}_- (R_{\dot m}-1) +{\dot \mu}_+
\eqno{(17b)}
$$
Here, $M_{disk}$ is the instantaneous mass of the disk.
Since $R_{\dot m} >1$, the disk has to evacuate. These cases hint that the assumptions of the
steady solution breaks down completely and the solutions may become become highly time dependent.

\subsection{Isothermal Outflow}

\noindent We assume that\\
\noindent (a) The outflow has exactly the {\it same} temperature as that of the
post-shock flow, but the energy is not conserved as matter goes from disk to the
jet. In other words the outflow is kept in a thermal bath of temperature as that of the
post-shock flow. 

\noindent (b)  Same as (b) of \S 3.1.

\noindent (c) The post-shock proton temperature is determined from the inflow accretion rate
${\dot M}_{in}$ using the consideration of Comptonization of the advective region. 
The procedure to compute typical proton temperature as a function of the incoming accretion rate
has been adopted from Chakrabarti (1997b). 

\noindent (d) The polytropic index of the inflow can be varied but that of the outflow is
always unity.

Thus a supply of parameters ${\cal E}$, $\lambda$ and $\gamma$ makes
a self-consistent computation of $R_{\dot m}$ possible when the shock is present. When the shock is absent,
the compression ratio of the gas at the pressure maximum between the inflow and the outflow
$R_{comp}$ is supplied as a free parameter exactly as in the polytropic case.

\noindent The following procedure is adopted to obtain a complete solution:\\

\noindent (a) From eqs. (9) and (10) we derive an expression for the derivative,
$$
{\frac{d{\vartheta}}{dr}|_{iso}}=\frac{\displaystyle{{\frac{{C_s}^2}{{\cal A}(r)}}
{\frac{d{\cal A}(r)}{dr}}+
{\frac{\lambda^2}{{r_m}^3(r)}}
{\frac{d{{r_m}(r)}}{dr}}-
{\frac{1}{2{({r_c}-1)^2}}}}}
{\displaystyle{\vartheta_{iso}-{\frac{{C_s}^2}{\vartheta_{iso}}}}} .
\eqno{(18)}
$$
At the sonic point, the numerator and denominator separately vanish, and give rise to the
so-called sonic point conditions:
$$
{\frac{{C_s}^2}{{{\cal A}_{co}}(r)}} {{\frac{d{{\cal A}(r)}}{dr}} |_{co}} +{\frac{\lambda^2}{{{r_{m_{co}}}^3}(r)}}
{{\frac{d{{r_m}(r)}}{dr}}|_{co}} -{\frac{1}{2{{({r_{co}}-1)}^2}}} = 0 ,
\eqno{(19a)}
$$
and
$$
{{\vartheta}_{co}}=C_s ,
\eqno{(19b)}
$$
where, the subscript $co$ represents the quantities at the sonic point of the outflow. The derivative
of the flow at the sonic point is computed using the L'Hospital's rule.  The procedure is otherwise
similar to those mentioned in the polytropic case and we do not repeat them here.

\section{Results}

\subsection{Polytropic outflow coming from the post-shock accretion disk}

Figure 1 shows a typical solution which combines the accretion and the outflow. The input parameters
are ${\cal E}=0.0005$, ${\lambda=1.75}$ and $\gamma=4/3$ corresponding to 
relativistic inflow. The solid curve with an arrow represents 
the pre-shock region of the inflow and the long-dashed curve represents the post-shock
inflow which enters the black hole after passing through the inner sonic point (I).
The solid vertical line at $X_{s3}$ (the leftmost vertical transition) with 
double arrow represents the shock transition obtained 
with exact Rankine-Hugoniot condition (i.e., with no mass loss). 
The actual shock location obtained with modified Rankine-Hugoniot condition
(eq. 15) is farther out from the original location $X_{s3}$.
Three vertical lines connected with the corresponding dotted curves represent 
three outflow solutions for the parameters $\gamma_{o}=1.3$ 
(top), $1.15$ (middle) and $1.05$ (bottom). The outflow
branches shown pass through the corresponding sonic points. It is
evident from the figure that the outflow  moves along solution curves completely different from that
of the `wind solution' of the inflow which passes through the outer sonic point `O'.
The mass loss ratio $R_{\dot m}$ in these cases are $0.256$, $0.159$ and $0.085$ 
respectively. Figure 2 shows the ratio $R_{\dot m}$ as $\gamma_{o}$ is varied. Only the 
range of $\gamma_{o}$ and energy for which the shock-solution is present is shown here. 
The general conclusion is that as $\gamma_{o}$ is increased  the ratio 
is also increased non-linearly. When the inflow rate is
very low, due to paucity of the photons, the outflow is not heated very much 
and $\gamma_{o}$ remains higher. The reverse is true when the  accretion rate is higher.
Thus, effectively, the ratio $R_{\dot m}$ is going up with the decrease in 
${\dot M}_{in}$. In passing we remark that with the variation in the inflow
angular momentum, $\lambda$, the result does not change significantly,
and $R_{\dot m}$ changes only by a couple of percentage at the most. 

In Fig. 3a, we show the variation of the ratio $R_{\dot m}$ of the mass outflow 
rate and inflow rate as a function of the shock-strength (dotted) $M_-/M_+$ 
(Here, $M_-$ and $M_+$ are the Mach numbers of the pre- and post-shock flows 
respectively.), the compression ratio  (solid) $\Sigma_+/\Sigma_-$ 
(Here, $\Sigma_-$ and $\Sigma_+$ are the vertically integrated matter 
densities in the pre- and post- shock flows respectively), and 
the stable shock location (dashed) $X_{s3}$ (in the notation of C89).
Other parameters are $\lambda=1.75$ and $\gamma_{o} = 1.05$. 
Note that the ratio $R_{\dot m}$ does not peak near the strongest 
shocks! Shocks are stronger when they are located closer to the black 
hole, i.e., for smaller energies. It is interesting to note that the general shape of $R_{\dot m}$ 
variation roughly agrees with analytical results of Chakrabarti (1997a; 1998) where
it was shown that for a finite outflow compression ratio must be non-zero
and  $R_{\dot m}$ peaks at an intermediate compression ratio. The non-monotonic behavior 
is more clearly seen in lowest curve of Fig. 3b where $R_{\dot m}$ is plotted 
as a function of the specific energy ${\cal E}$ (along x-axis) 
and $\gamma_{o}$ (marked on each curve). Specific angular momentum 
is chosen to be $\lambda=1.75$ as before. The tendency of the peak in $R_{\dot m}$ 
is primarily because as ${\cal E}$ is increased, the shock location 
is increased which generally increases the outflowing area ${\cal A}(r)$ 
at the shock location. However, the density of the outflow at the shock, 
as well as the velocity of the outflow at the shock increases.  The outflow
rate, which is a product of these quantities, thus shows a peak. 
For the sake of comparison, we present the results for $\gamma_o=1.05$ (dashed curve)
when the Rankine-Hugoniot relation was not corrected by eq. (15). The result
generally remains the same because of two competing effects: decrease in 
post-shock density and increase in the area from the the outflow is launched
(i.e., area between the black hole and the shock) as well as the launching 
velocity of the jet at the shock.

To have a better insight of the behavior of the outflow we plot in Fig. 4 $R_{\dot m}$
as a function of the polytropic index of the {\it incoming} flow 
($\gamma$) for $\gamma_o=1.1$, ${\cal E}=0.002$ and $\lambda=1.75$. The range 
of $\gamma$ shown is the range for which shock forms in the flow. We also plot the
variation of injection velocity $\vartheta_{inj}$, injection density $\rho_{inj}$  
and area ${\cal A}(r)$ of the outflow at the location where the outflow leaves the disk.
The incoming accretion rate has been chosen to be $0.3$ (in units of the Eddington rate).
These quantities are scaled from the corresponding dimensionless 
quantities as $\vartheta_{inj} \rightarrow 0.1\vartheta_{inj} $, 
$\rho_{inj} \rightarrow 10^{22} \rho_{inj}$ and ${\cal A} 
\rightarrow 10^{-4} {\cal A} $ respectively in order to bring them in the
same scale. With the increase in $\gamma$, the shock location is increased, and therefore the
cross-sectional area of the outflow goes up. The injection velocity goes up
(albeit very slowly) as the shock recedes, since the injection surface
(CENBOL) comes closer to the outflow sonic point. However, the density goes down
(gas is less denser). This anti-correlation is reflected in the
peak of $R_{\dot m}$.

So far, we assumed that the specific angular momentum of the outflow
is exactly the same as that of the inflow, while in reality it could be
different due to presence of viscosity. In the outflow, a major 
source of viscosity is the radiative viscosity whose coefficient is,
$$
\eta = \frac{4 a T^4}{15 \kappa_T c \rho}  \  {\rm cm^2}\ {\rm sec^{-1}}
\eqno{(20)}
$$
This could be significant, since the temperature of the outflow is high, but the density is low.
Assuming that the angular momentum distribution reaches a steady state
inside the jet (Chakrabarti, 1986 and references therein),
$$
l_j= C_j R^{n_j}
\eqno{(21)}
$$
where $C_j$ and $n_j$ are constants, the vanishing condition of the azimuthal
velocity on the axis requires that $n_j>1$ inside the jet. 
The matter distribution in the {\it rotationally dominant}
region of the `pre-jet' is computed by integrating Euler 
equation. It is easy to show that the `hollow' jet thus 
produced carry most of the matter and angular momentum
in the outer layers of the jet (Chakrabarti, 1986). In other words, 
the average angular momentum of the outflow {\it away from the base} 
may remain roughly constant even in presence of viscosity. This is 
to be contrasted with the disk, where matter is more dense towards the 
centre while more angular momentum is concentrated towards the outer edge.
If, however, the average angular momentum {\it at the base} of the
outflow goes does down due to losses to ambient medium, by, say, a factor of two, 
we find that the mass loss rate is also reduced by around the same factor. 
This shows that the outflow is at least partially centrifugally driven.

An important point to note: the ratio between the `specific 
entropy measure' of the outflow
to that of the post-shock inflow is obtained from the 
definitions of entropy accretion rate ${\dot {\cal M}}$ 
(see, eq. 3b):
$$
\frac{K_o}{K_+} = \frac{\dot{\cal M}_{out}^{\gamma_o-1}}{\dot{\cal M}_{+}^{\gamma-1}}
\left (\frac{1-R_{\dot m}}{R_{\dot m}}\right)^{\gamma_o-1} {\dot M}_+^{({\gamma-\gamma_o})}
\frac{\gamma}{\gamma_o}
\eqno{(22)}
$$
As $R_{\dot m} \rightarrow 1$, $\frac{K_o}{K_+} \rightarrow 0$. Thus, we expect that
for a polytropic flow with shocks, a hundred percent outflow is impossible since the outgoing
entropy must be higher. In isothermal outflows such simple consideration do not apply.

If we introduce an extra radiation pressure term (with a term like
$\Gamma/r^2$ in the radial force equation, where $\Gamma$ is the contribution
due to radiative process), particularly important for 
neutron stars, the outcome is significant. In the inflow, outward radiation
pressure weakens gravity and thus the shock is located farther out. The temperature is
cooler and therefore the outflow rate is lower. If the term is introduced only in the
outflow, the effect is not significant.

\subsection{Polytropic outflow coming from the region of the maximum pressure}

In this case, the inflow parameters are chosen from region {\bf I} (see C96c)
so that the shocks do not form. Here, the inflow passes through 
the inner sonic point only. The outflow is assumed to be originated
from the regions where the polytropic inflow has a maximum pressure. This assumption is
justified, since it is expected that winds would get the maximum kick at this region.  
Figure 5a shows a typical solution. The arrowed solid curve shows the inflow 
and the dotted arrowed curves show the outflows 
for $\gamma_o=1.3$ (top), $1.1$ (middle) and $1.01$ (bottom). 
The ratio $R_{\dot m}$ in these cases is given by $0.66$, $0.30$ and $0.09$ respectively. 
The specific energy and angular momentum are chosen to be ${\cal E}=0.00584$ and 
$\lambda=1.8145$ respectively. The pressure maximum occurs outside the inner sonic point at $r_p$ when 
the flow is still subsonic. Figure 5b shows the variation of 
thermal pressure of the flow with radial distance. The peak is clearly visible. Since the
pressure maximum occurs very close to the black hole as compared to the location of the
shock, the area of the outflow is smaller, but the radial velocity as well as the density of
matter at the base of the outflow are much higher. As a result 
the outflow rate is exorbitantly higher compared to the shock case. Figure 6 shows the 
ratio $R_{\dot m}$ as a function of $\gamma_{o}$ for various choices of the
compression ratio $R_{comp}$ of the outflowing gas at the pressure maximum: $R_{comp}=2$ for the
rightmost curve and $7$ for the leftmost curve. We have purposely removed the solutions
with $R_{\dot m} >1$, because the solution should be inherently time-dependent (see, eq. 17b)
in these cases and a steady state approach is not supposed to be trusted completely.
This is different from the results
of \S 4.1, where shocks are considered, since $R_{\dot m}$ is non-monotonic in that case.
It is also found that in some range of parameters, the very high massflow   
could take place even for smaller compression ratios especially
when the sonic point of the outflow $r_{o}$ is right outside the pressure maximum. These cases
can cause runaway instabilities by rapidly evacuating the disk. They may be responsible for
quiescent states in X-ray novae systems and also in some systems with massive black holes 
(e.g., our own galactic centre?), although it is not clear if such a phase of profuse mass loss
has been directly observed.

The location of maximum pressure being close to the black hole, it  may be generally
very difficult to generate the outflow from this region. Thus, it is expected that the
ratio $R_{\dot m}$ would be larger when the maximum pressure is located farther out.
This is exactly what we see in Fig. 7, where we plot $R_{\dot m}$ against the location of
the pressure maximum (solid curve). Secondly, if our guess that the outflow 
rate could be related to the pressure is correct, then the rate 
should increase as the pressure at the maximum rises. That is also 
observed in Fig. 7. We plot here $R_{\dot m}$ as a function of the 
actual pressure at the pressure maximum (dotted curve). The mass loss is found to be a strongly
correlated with the thermal pressure. Here we have multiplied non-dimensional thermal 
pressure by $1.5 \times 10^{24}$ in order to bring it in the same scale.

\subsection{Isothermal outflow coming from the post-shock accretion disk}

In this case, the outflow is assumed to be isothermal. The temperature of the outflow
is obtained from the proton temperature of the advective region of the disk. The proton temperature
is obtained using the Comptonization, bremsstrahlung, inverse bremsstrahlung and Coulomb processes
(Chakrabarti, 1997b and references therein). Figure 8 shows the effective
proton temperature and the electron temperature of the post-shock advective region as a function of the
accretion rate (in units of Eddington rate, in logarithmic scale)  of the Keplerian component of the disk. The
diagram is drawn for a black hole of mass $10M_\odot$. The soft
X-ray luminosity for stellar mass black holes or the UV luminosity of massive black holes
is basically dictated by the Keplerian rate of the disk. It is clear that as the accretion rate of the
Keplerian disk is increased, the advective region gets cooler as is expected.

In Fig. 9a, we show the ratio $R_{\dot m}$ as a function of the accretion rate (in 
units of Eddington rate) of the incoming flow for a range of the specific angular momentum. 
In the low luminosity objects the ratio is larger. Angular momentum is varied 
from $\lambda=1.7$ (top curve), $1.725$ (middle curve) and $1.75$ (bottom curve). 
The specific energy is ${\cal E}=0.003$. Here we have used the modified Rankine-Hugoniot 
relation as before (eq. 15). The ratio  $R_{\dot m}$ is clearly very 
sensitive to the angular momentum since it changes the shock location 
rapidly and therefore changes the post-shock temperature very much. 
We also plot the outflux of angular momentum $F ({\lambda})=\lambda {\dot m}_{in} R_{\dot m}$
which has a maximum at intermediate accretion rates. In dimensional units, these
quantities represent significant fractions of angular momentum  of the entire disk
and therefore the rotating outflow can help accretion processes. Curves are drawn for 
different $\lambda$ as above. In Fig. 9b, we plot the variation of the ratio directly 
with the proton temperature of the advecting region. The outflow is clearly thermally 
driven. Hotter flow produces more winds as is expected. The angular momentum 
associated with each curve is same as before.  

\subsection{Isothermal outflow coming from the region of the maximum pressure}

This case produces very similar result as in the above case, except that like Section 4.2 the
outflow rate becomes very close to a hundred percent of the inflow
rate when the proton temperature is very high. Thus, when the accretion rate of the 
Keplerian flow is very small, the outflow rate becomes very high, close to evacuating the
disk. As noted before, this may also be related to the quiescent state of the X-ray novae.

\section{Conclusions}

In this paper, we have computed the mass outflow rate from the advective accretion disks
around galactic and extra-galactic black holes. Since the general physics
of advective flows are similar around a neutron star, we believe that the
conclusions may remain roughly similar provided the shock at $X_{s3}$ forms, 
although the boundary layer (which corresponds to $X_{s1}$ in C89 notation; also
see C96b) of the neutron star, where half of the binding energy could be released,
may be more luminous than that of a black hole and may thus 
affect the outflow rate. We have chosen a limited number of free parameters 
just sufficient to describe the inflow and only one extra parameter for 
the outflow. We find that the outflow rates can vary from a very few percentage 
of the inflow rate, to as much as the inflow rate (causing 
almost complete evacuation of the accretion disk) depending on the inflow parameters.
For the  first time, it became possible to use the exact transonic solutions 
for both the disks and the winds and combine them to form a self-consistent 
disk-outflow system. Although we present results when centrifugally supported
boundary layers are considered around a black hole, it is evident that the
result is general, namely, if such a barrier is produced by other means, such as 
pre-heating (Chang \& Ostriker, 1985) or by pair-plasma pressure (Kazanas \& Ellison, 1986), 
outflows would also be produced (Das, in preparation).

The basic conclusions of this paper are the followings:\\

\noindent a) It is possible that most of the outflows are coming from the centrifugally 
supported boundary layer (CENBOL) of the accretion disks.\\
\noindent b) The outflow rate generally increases with the proton temperature of CENBOL. In other
words, winds are, at least partially, thermally driven. This is reflected more strongly 
when the outflow is isothermal.\\
\noindent c) Even though specific angular momentum of the flow increases the size of the CENBOL,
and one would have expected a higher mass flux in the wind, we find that the rate of the
outflow is actually anti-correlated with the $\lambda$ of the inflow. On the other hand, if 
the angular momentum of the outflow is reduced by hand, we find that the rate of the outflow is correlated with  
$\lambda$ of the outflow. This suggests that the outflow is partially centrifugally driven as well.\\
\noindent d) The ratio $R_{\dot m}$ is generally anti-correlated with the inflow accretion rate. That is,
disks of lower luminosity would produce higher $R_{\dot m}$.\\
\noindent e) Generally speaking, supersonic region of the inflow do not have pressure maxima. Thus,
outflows emerge from the subsonic region of the inflow, whether the shock actually forms or not.\\

In this paper, we assumed that the magnetic field is absent. Centrifugally
driven winds (e.g., Blandford \& Payne, 1982), electromagnetically accelerated
winds (Lovelace, 1976; Lovelace, Wang \& Sulkanen, 1987; Contopoulos \& Lovelace,
1994 etc.) and winds from Blandford \& Znajek (1977) process have not been 
considered here. These effects have so far been considered in the context of
a Keplerian disk and {\it not} in the context of sub-Keplerian flows on which 
we concentrate here. Secondly, whereas the entire Keplerian disk was assumed to participate
in wind formation, here we suggest that CENBOL is the major source of outflowing matter.
It is not unreasonable to assume that CENBOL would still form when magnetic fields are present (Chakrabarti,
1990) and since the Alfv\'en speed is, by definition, higher compared to the
sound speed, the acceleration, and therefore the mass outflow would also be higher
than what we computed here. Such works would be carried out in future.

In the literature, not many results are present which deal with exact computations of the mass outflow
rate. Molteni, Lanzafame \& Chakrabarti (1994), in their SPH simulations, found that the
ratio could be as high as $15-20$ per cent when the flow is steady. In Ryu, 
Chakrabarti \& Molteni (1997),  $10-15$ per cent of the steady outflow is seen 
and occasionally, even $150$ of the inflow is found to be ejected 
in non-stationary cases. Our result shows that high outflow rate is also possible, 
especially for absence of shocks and low luminosities. In Eggum, Coroniti \& Katz (1985), 
radiation dominated flows showed $R_{\dot m} \sim 0.004$, which also agrees with our 
results when we consider high accretion rates (see, e.g., Fig. 9a). 
Observationally, it is very difficult to obtain the outflow rate from a real system, as it depends on
too many uncertainties, such as filling factors and projection effects. In any case, with a
general knowledge of the outflow rate, we can now proceed to estimate several important quantities.
For example, it had been argued that  the composition of the disk changes due to nucleosynthesis
in accretion disks around black holes and this modified isotopes are deposited in the 
surroundings by outflows from the disks (Hogan \& Applegate 1987; Mukhopadhyay \& Chakrabarti, 1998
and references therein). Similarly, it is argued that outflows deposit magnetic flux tubes from accretion disks
into  the surroundings (Daly \& Loeb, 1990). Thus a knowledge of outflows are essential in understanding
varied physical phenomena in galactic environments.

A number of possible improvements could be made on this basic work. For instance, the 
effect of radiation pressure on both the inflows and outflows are to be taken into account
self-consistently, particularly when the neutron star winds are considered. 
A preliminary investigation with a $\Gamma/r^2$ force term (whose
effect is to weaken gravity) suggests that for non-zero $\Gamma$ in the inflow, the
mass-loss rate changes significantly. This is because the shock location increases
when $\Gamma$ is increased. This in turn reduces the mass loss rate. On the other hand,
the when $\Gamma$ is non-zero in the outflow, the effect is not very high, since the
outflow rate is generally driven by thermal effect of the {\it disk}
and not the wind. Similarly, we see a significant reduction of the outflow when the
average specific angular momentum of the outflow is reduced. This is expected since the
outflow is partially centrifugally driven. This effect is stronger when the outflow is isothermal.

An interesting situation arises when the polytropic index of the outflow is large 
and the compression ratio of the flow is also very high. In this case, the flow virtually
bounces back as the winds and the outflow rate can be equal to the
inflow rate or even higher, thereby evacuating the disk. In this range of parameters, most, if not all,
of our assumptions may breakdown completely because the situation could become inherently time-dependent.
It is possible that some of the black hole systems, including that in our own galactic centre,
may have undergone such evacuation phase in the past and gone into quiescent phase. 

So far, we made the computations around a Schwarzschild black hole. In the case of a Kerr black hole
(C96c), the shock locations will come closer and the outflow rates should become higher. Similarly
magnetic field will change the sonic point locations significantly (C90). The mass outflow
rates in these conditions are being studied and the results would be reported elsewhere (Das, 1998b).
We made a few assumptions, some of which may be questionable. Our assumption of the 
wind formation from the post-shock region is very similar to the assumptions involved in 
computing mass outflow rate from stellar atmospheres. This is because the post-shock
region behaves like a boundary layer. However, when the shocks
are not formed, we assumed that the outflow comes out of the region of maximum pressure.
Whereas, this may be a plausible assumption, it is not clear if strong outflows
actually form from that close to the black hole. Similarly, whereas the assumption of
isothermality in the stellar system may be justified, it is questionable if this can be rigorously
valid in the present system since the proton temperatures are too
high, and to maintain that high temperature till the sonic point, the disk has to deposit 
enormous energy into the winds. Nevertheless, we believe that our calculation 
is sufficiently illustrative and gives a direction which can be followed in the future 
for further refinements.

We thank the unknown referee for his helpful suggestions. 

{}

\newpage

\centerline {Figure Captions}

\noindent Fig. 1: Few typical solutions which combine accretion and outflow. Input parameters
are ${\cal E}=0.0005$, ${\lambda=1.75}$ and $\gamma=4/3$. 
Solid curve with an incoming arrow represents 
the pre-shock region of the inflow and the dashed curve with an incoming arrow
represents post-shock inflow which enters the black hole after passing through the inner sonic point
(I). Dotted curves are the outflows for various $\gamma_o$ (marked). Open circles are sonic points
of the outflowing winds and the crossing point `O' is the outer sonic point of the inflow. The 
leftmost shock transition ($X_{s3}$) is obtained from unmodified Rankine-Hugoniot condition,
while  the other transitions are obtained when the mass-outflow is taken into account.

\noindent Fig. 2: Ratio $R_{\dot m}$ as outflowing polytropic index $\gamma_{o}$ is varied. 
Only the range of $\gamma_{o}$ for which the shock-solution is present is shown here. 
As $\gamma_{o}$ is increased, the ratio is also increased. Since $\gamma_o$ is 
generally anti-correlated with ${\dot m}_{in}$, 
this implies that $R_{\dot m}$ is correlated with ${\dot m}_{in}$.

\noindent Fig. 3: Variation of $R_{\dot m}$ as a function of (a) shock-strength $M_-/M_+$ (dotted)
compression ratio $\Sigma_+/\Sigma_-$ (solid) and the shock location $X_{s3}$ (dashed)
for $\gamma_{o} = 1.05$ and (b) specific energy ${\cal E}$ for various $\gamma_o$ (marked). 
$\lambda=1.75$ throughout. 

\noindent Fig. 4: $R_{\dot m}$ as a function of the polytropic index $\gamma$ of the inflow.
The range of $\gamma$ shown is the range for which shock forms in the flow. 
Suitably scaled density, velocity and area of the flow (at the base of the outflow)
on the disk surface are also shown. See text for details. Non-monotonicity 
in $R_{\dot m}$ can be understood by the fact that the shock location, i.e., 
the area ${\cal A} (r_{inj})$ and velocity $\vartheta_{inj}$ of the outflow {\it at the outflow origin} go up 
with $\gamma$, but the density $\rho_{inj}$ goes down.

\noindent Fig. 5a: Few typical solutions for outflows forming out of an advective disk
which does not include a standing shock wave.
Incoming arrowed solid curve shows the inflow and the dashed arrowed curves with outgoing
arrows show the outflows for $\gamma_o=1.3$ (top), $1.1$ (middle) and $1.01$ (bottom). 
At $r_p$, thermal pressure of the inflow is maximum.

\noindent Fig. 5b: Variation of thermal pressure $P_e$ of the incoming flow
with radial distance. In a shock-free hydrodynamic flow, winds may form
from the region around the pressure maximum.

\noindent Fig. 6: $R_{\dot m}$ as a function of outflowing polytropic index
$\gamma_{o}$ for various choices of the compression ratio $R_{comp}$ of the outflowing gas 
at the pressure maximum. From bottom to top curve, $R_{comp} = 2,\ 3,\ 4,\ 5,\ 6 $ \&  $7$
respectively. 

\noindent Fig. 7: Variation of $R_{\dot m}$ (solid) as a function of the location
$r_p$ of the maximum pressure and the non-dimensional pressure (dotted) $P_e(r_p)$
(multiplied by $1.5 \times 10^{24}$ to bring to scale) are plotted.

\noindent Fig. 8: Variation of effective proton (solid) and electron (dotted) 
temperatures in the advective regions of the accretion disk around a $10$ solar
mass black hole as functions of the accretion rates of the Keplerian flow. 
As Keplerian rate increases, both protons and elections cool down. 

\noindent Fig. 9(a-b): (a) Variation of $R_{\dot m}$ as functions of the Eddington rate of the
Keplerian component of the incoming flow for a range of the specific angular momentum. In the low luminosity
objects the ratio is larger. (b) $R_{\dot m}$ as a function of the proton temperature $T_p$ of the
post-shock region. In (a), $\lambda=1.7$ (top curve), $\lambda=1.725$ (middle) and $1.75$ (bottom curve).
In (a) the angular momentum flux $F(\lambda)$ of the outflow is also shown (dashed curve).

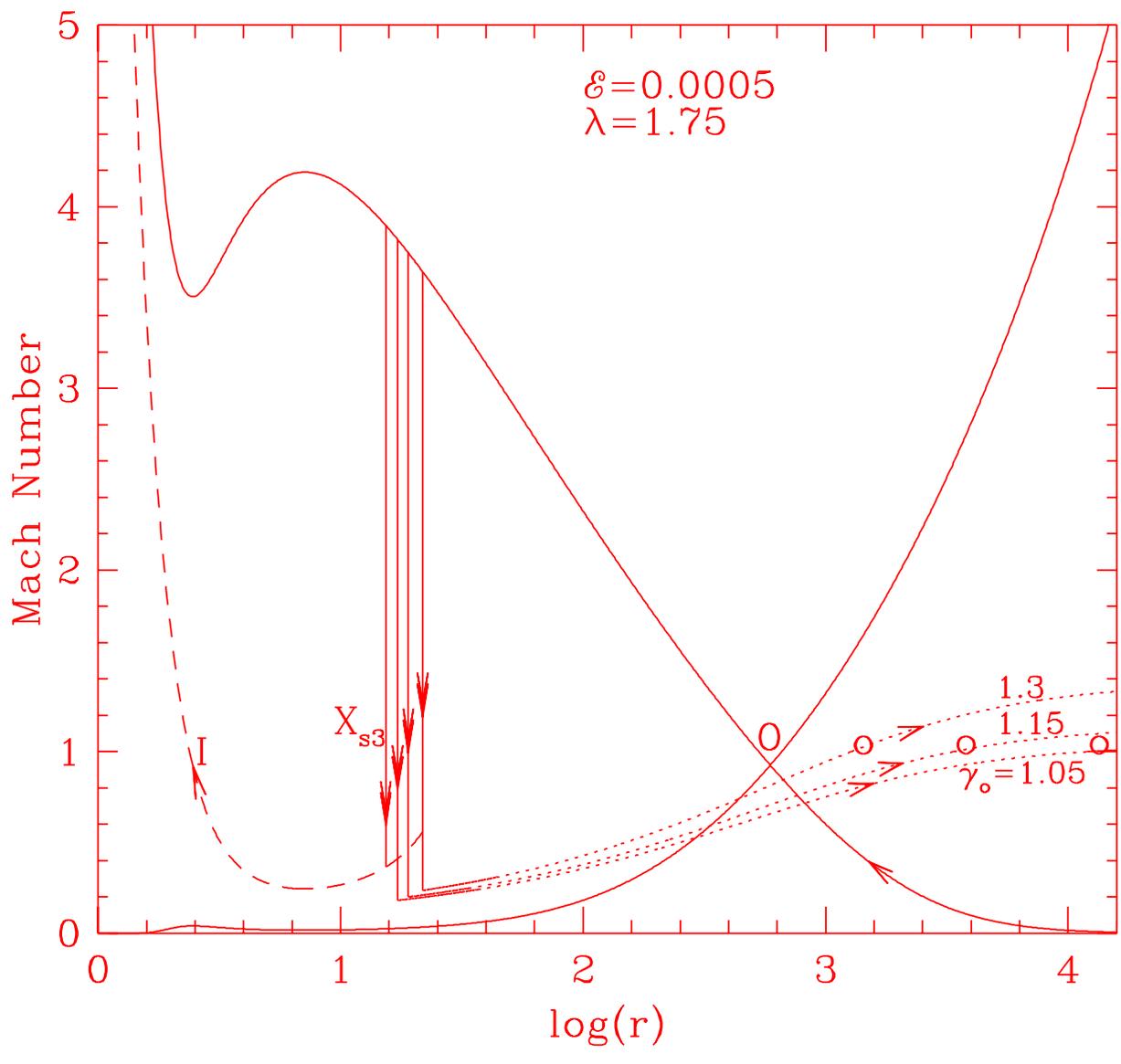
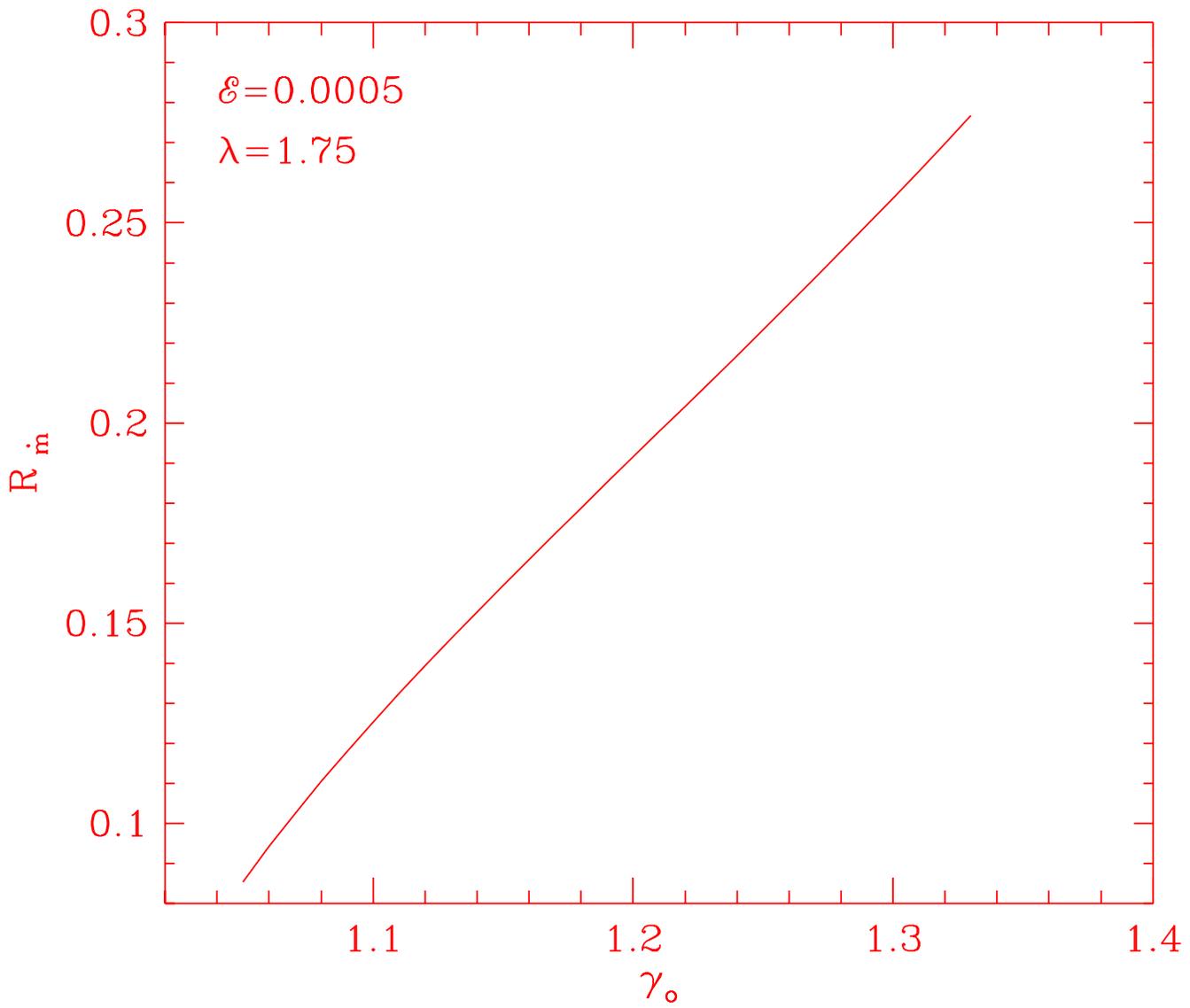
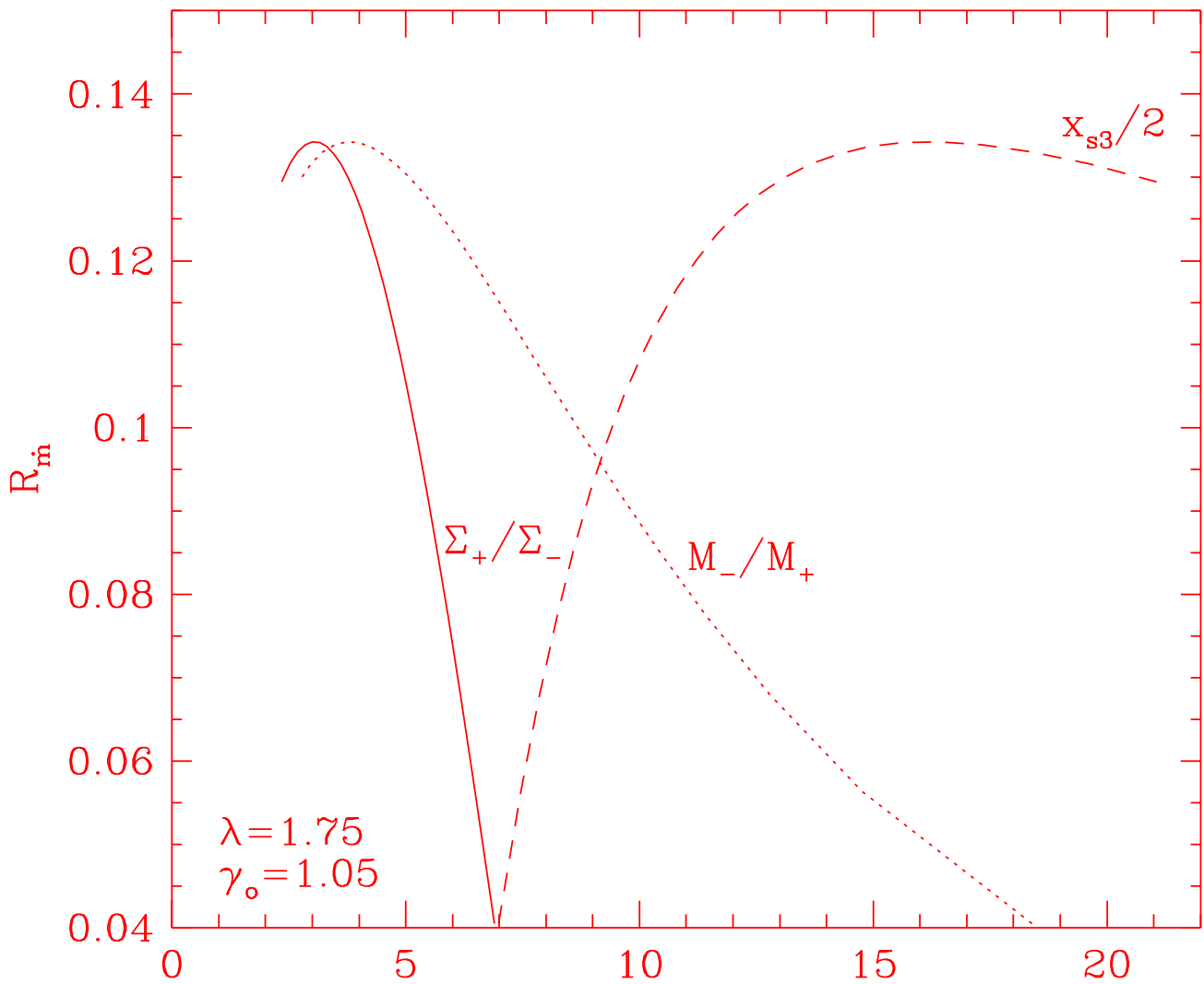
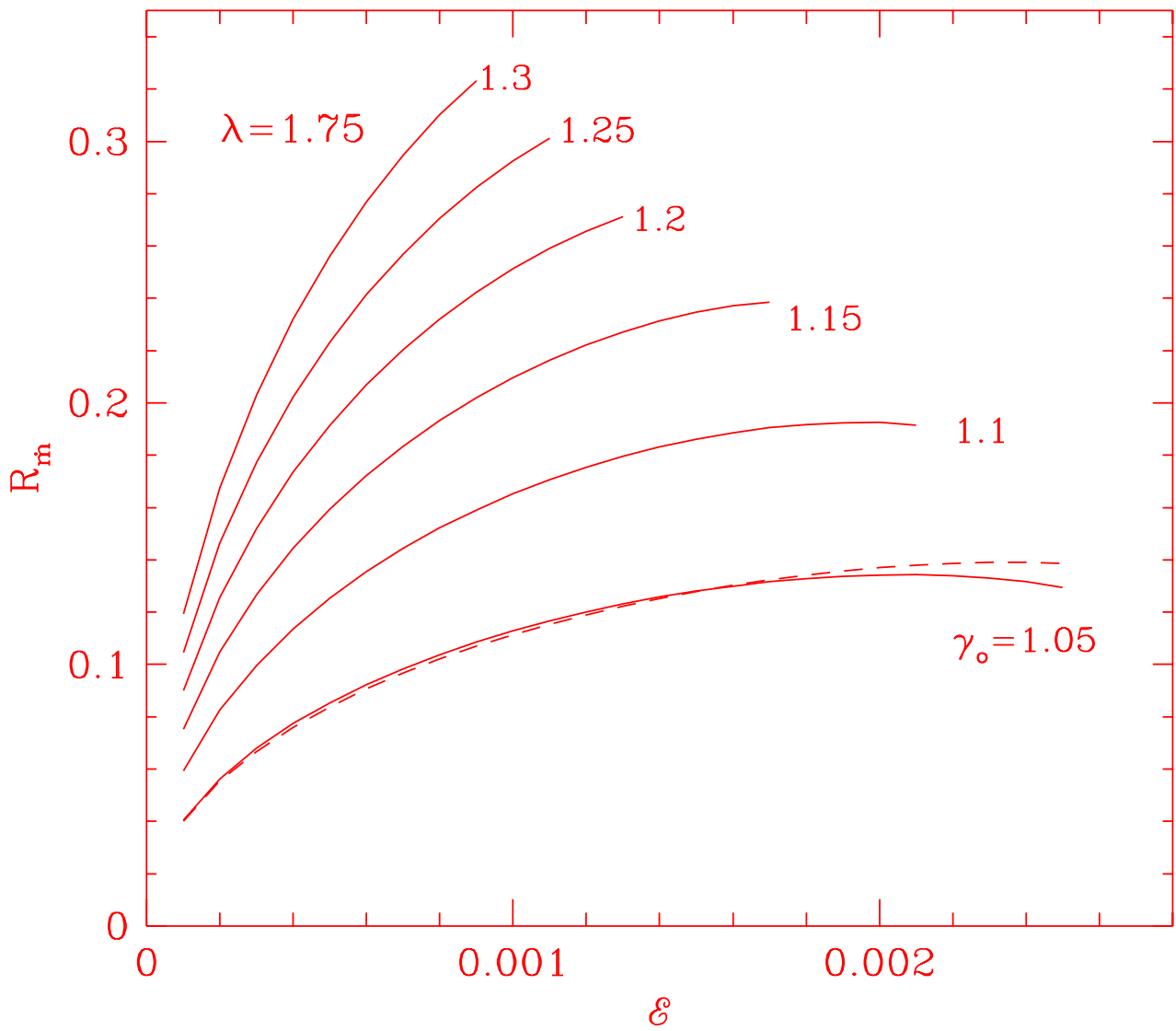
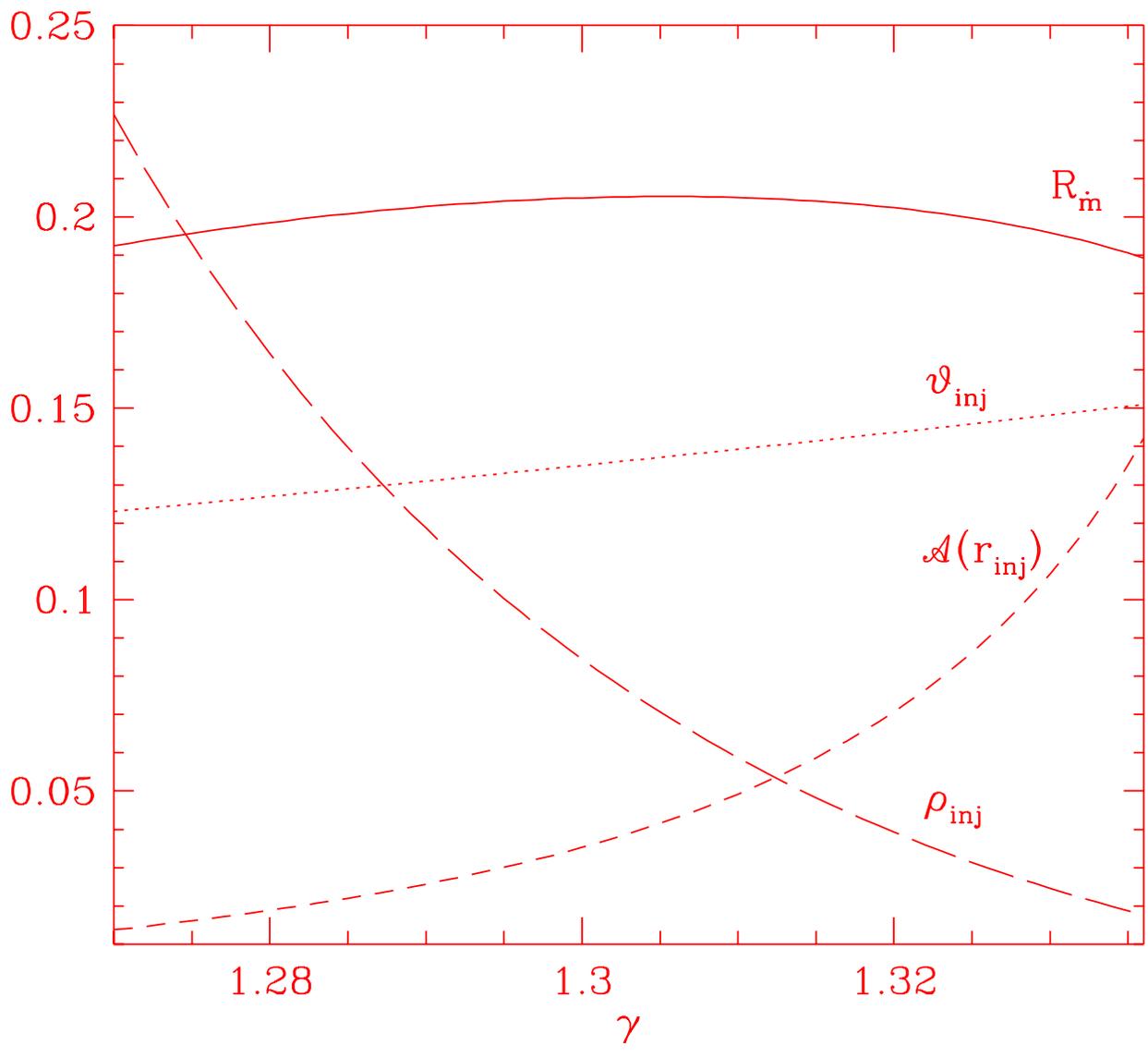
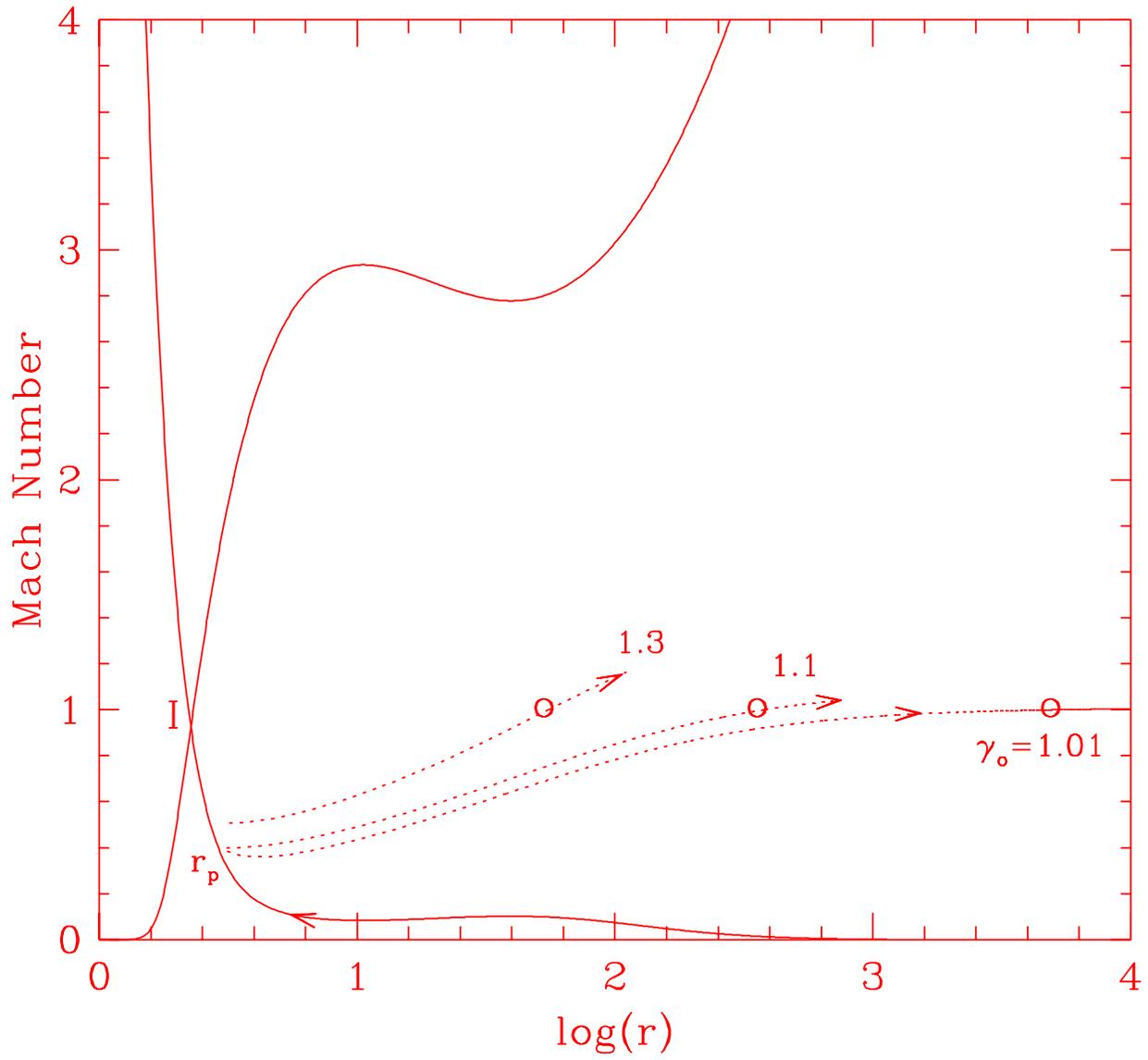
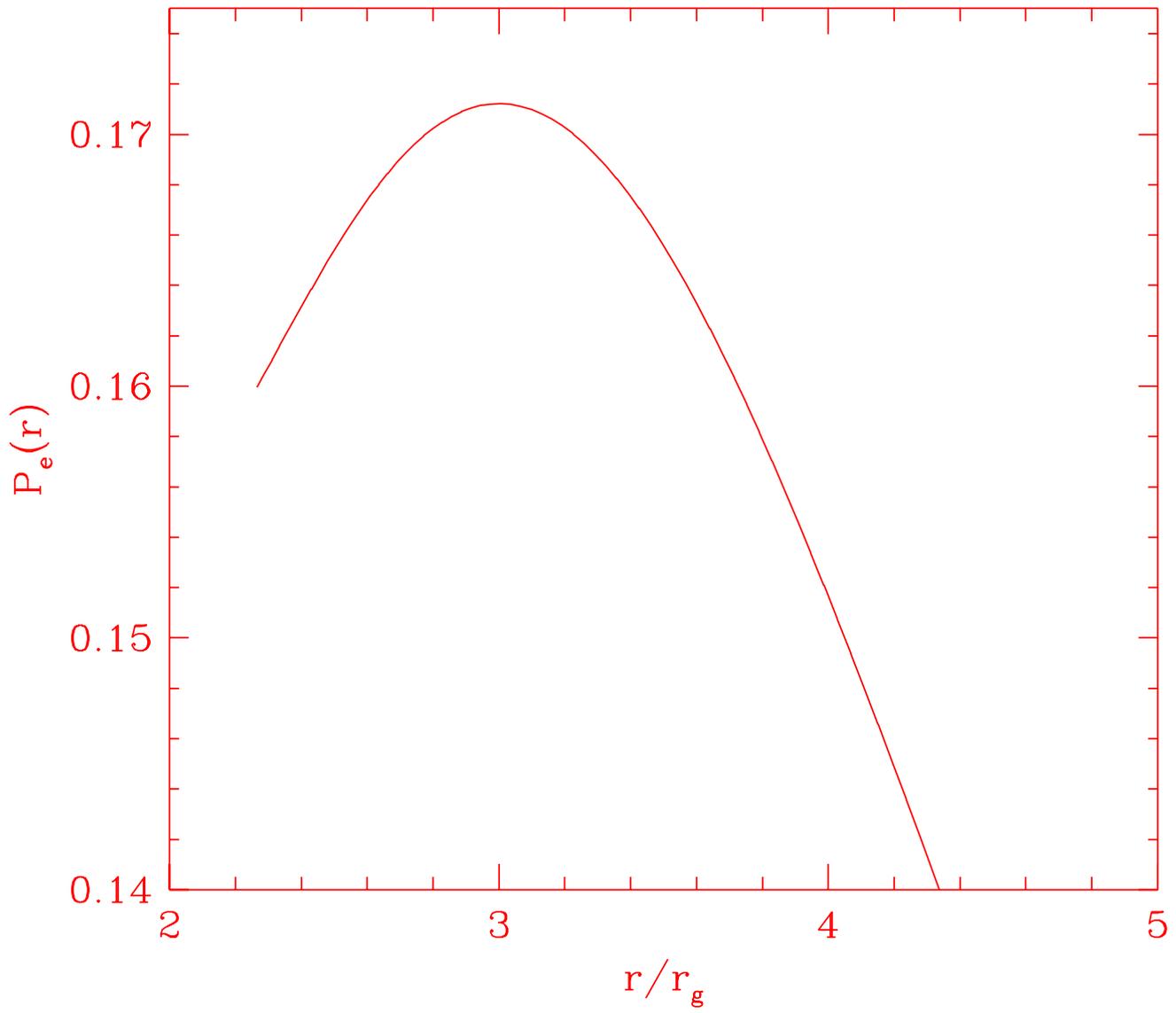
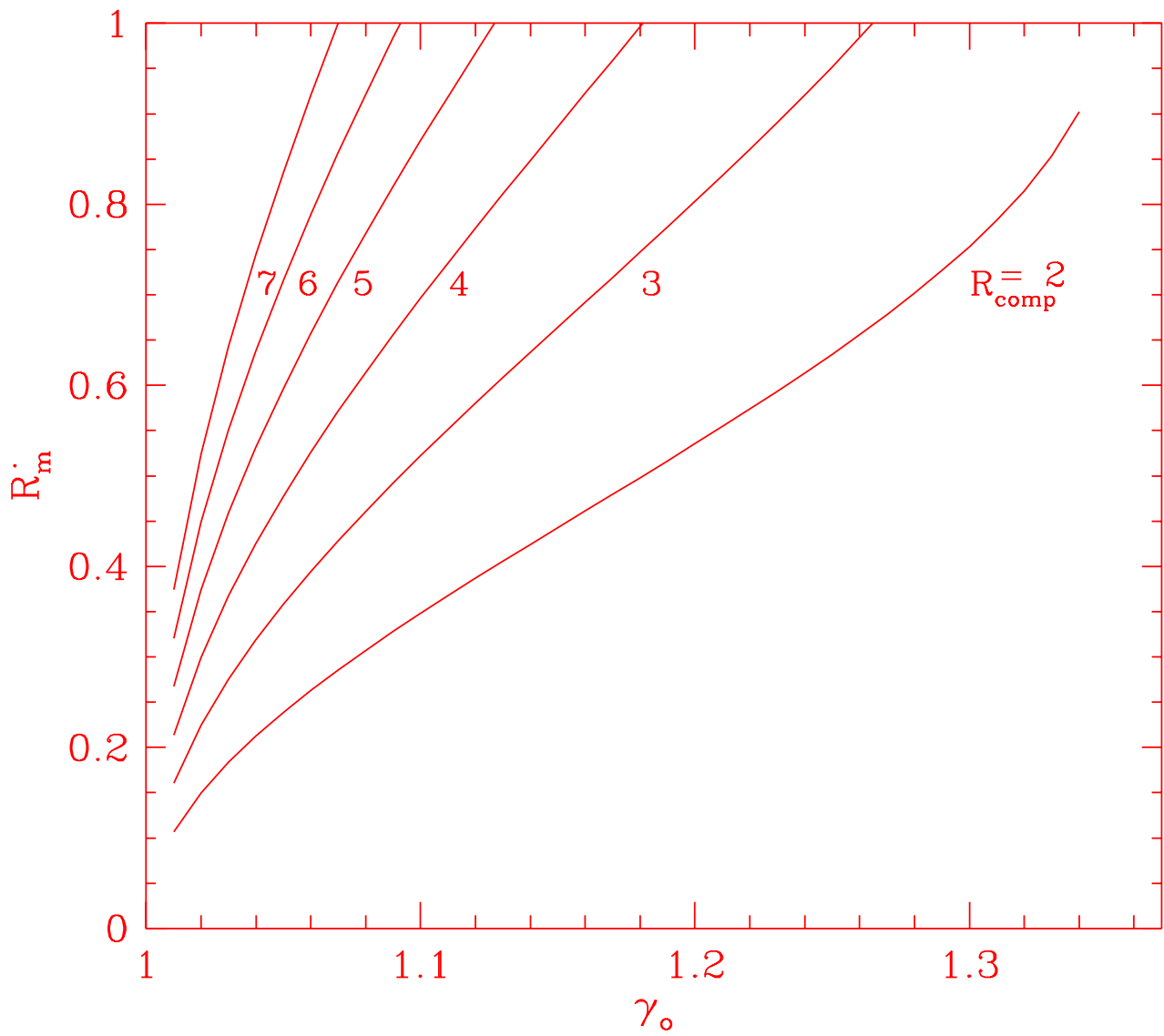
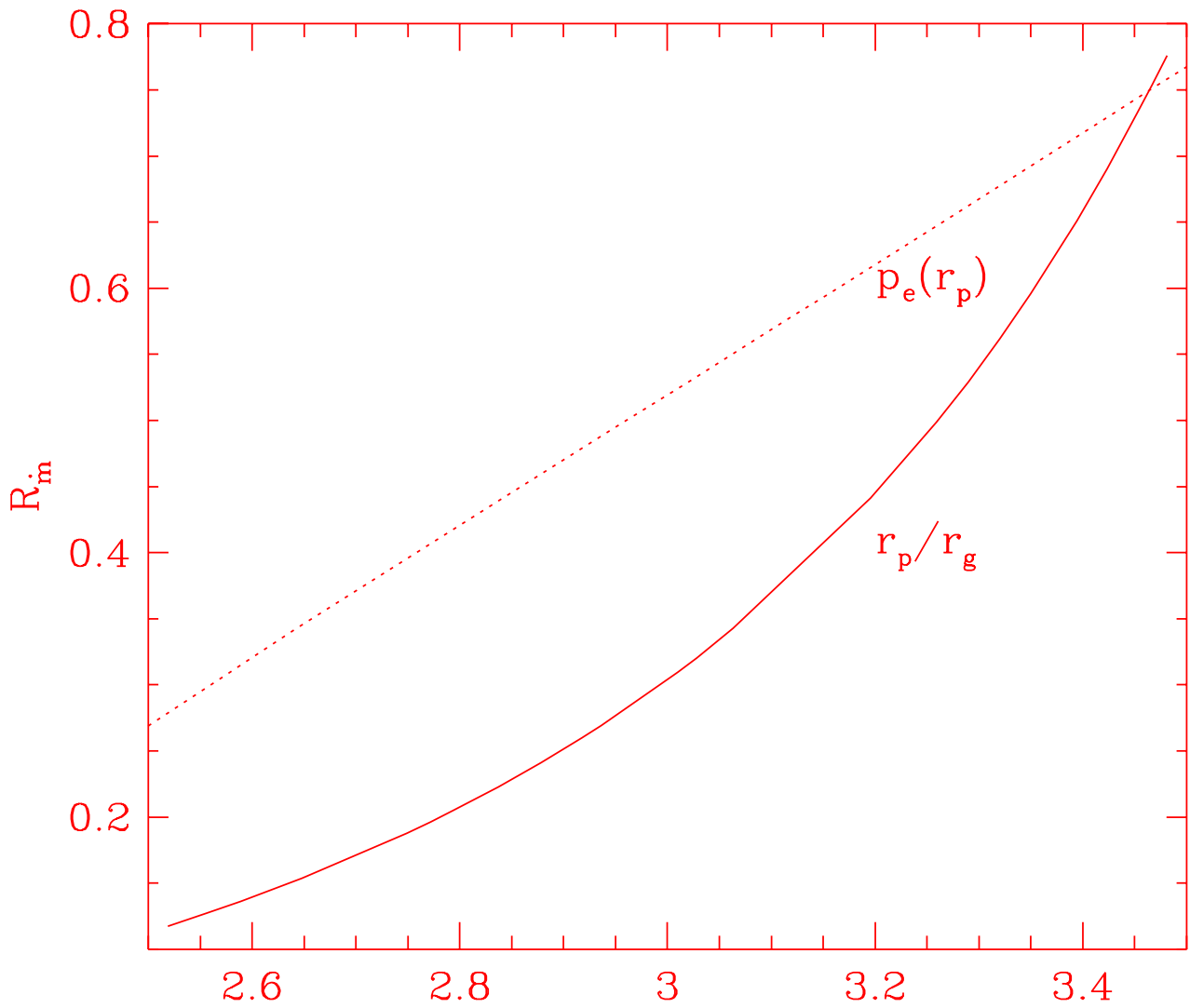
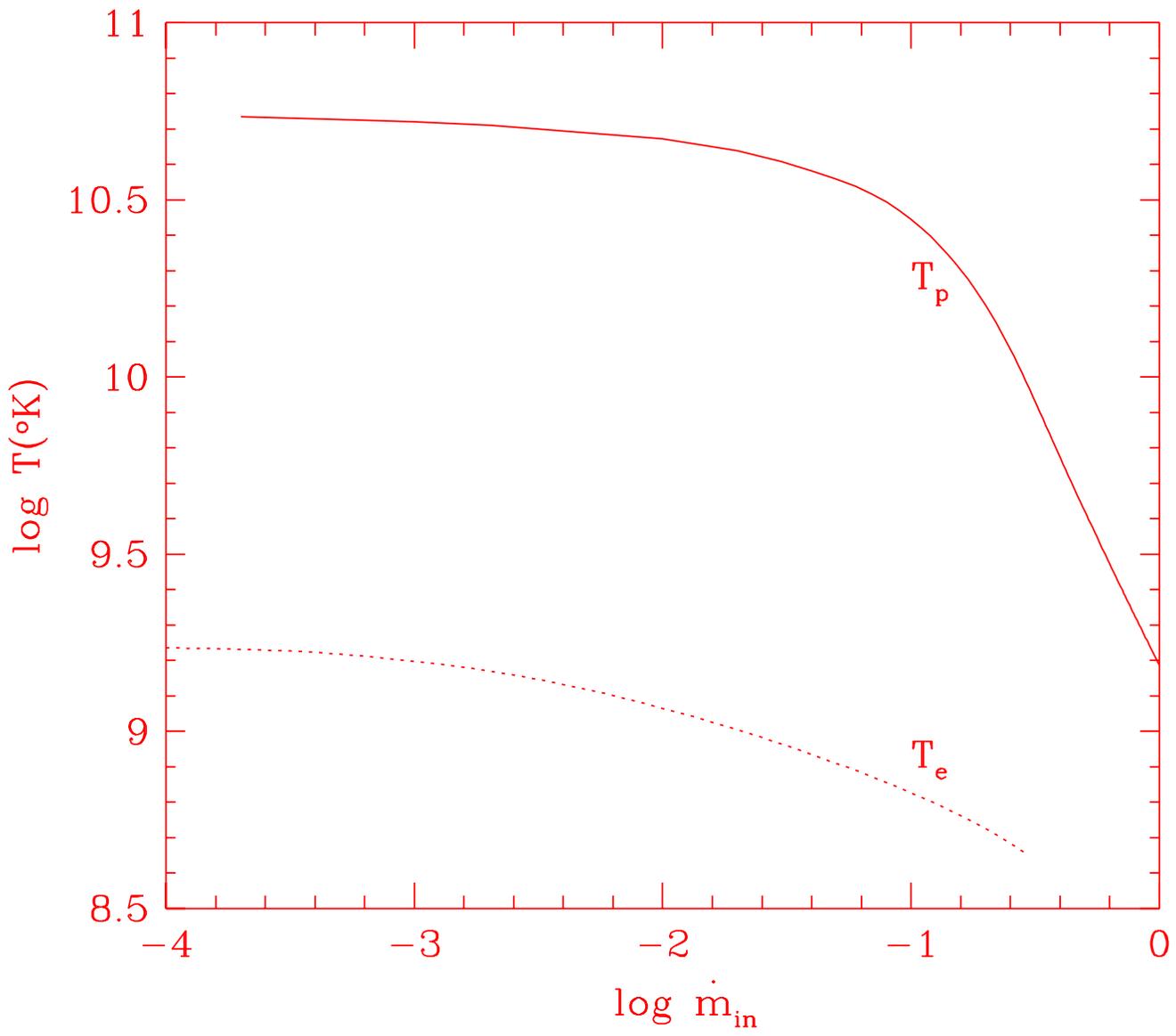
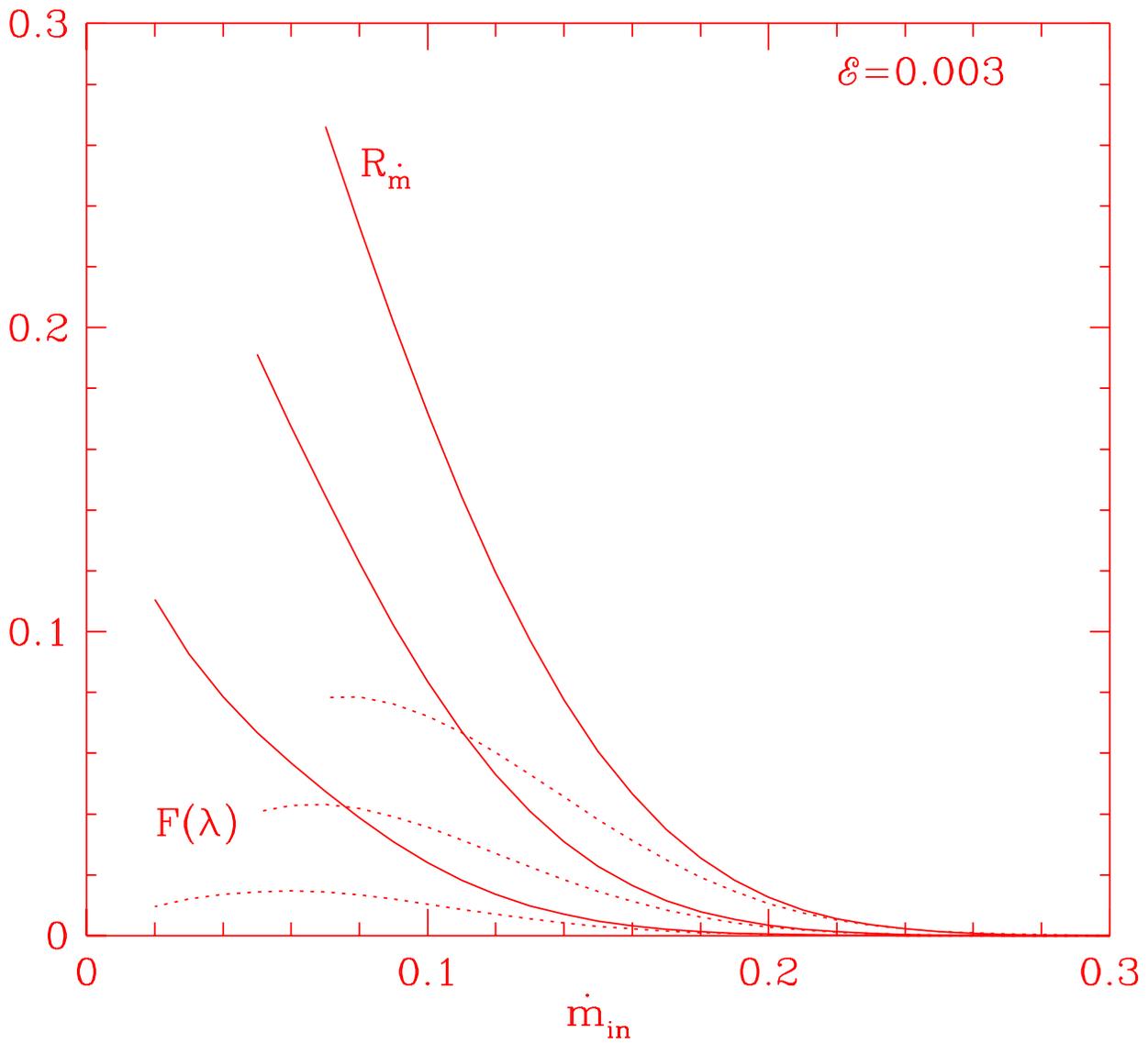
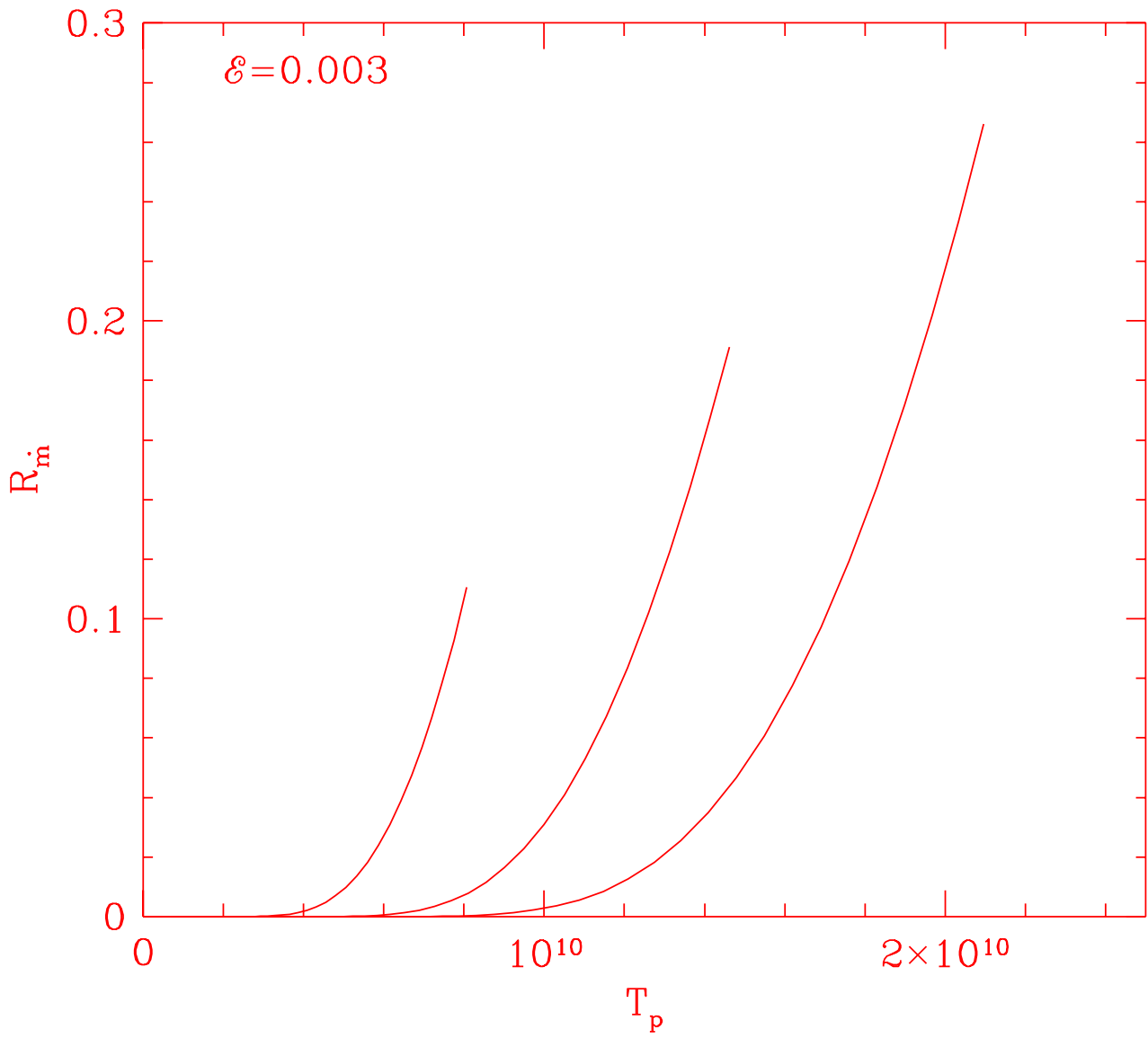
\end{document}